\documentclass[showpacs,a4paper,12pt,pra,onecolumn]{revtex4}
\usepackage{amsfonts}
\usepackage{amsmath}
\usepackage{amssymb}
\usepackage{graphicx}%
\providecommand{\U}[1]{\protect\rule{.1in}{.1in}}

\begin{document}
\title{Slicing the Fock space for state production and protection}
\author{R. F. Rossetti$^{1}$, G. D. de Moraes Neto$^{1}$, F. O. Prado$^{2}$, F. Brito$^{1}$,
and M. H. Y. Moussa$^{1}$}
\affiliation{$^{1}$Instituto de F\'{\i}sica de S\~{a}o Carlos, Universidade de S\~{a}o
Paulo, Caixa Postal 369, 13560-970, S\~{a}o Carlos, S\~{a}o Paulo, Brazil}
\affiliation{$^{2}$Universidade Federal do ABC, Rua Santa Ad\'{e}lia 166, Santo Andr\'{e},
S\~{a}o Paulo 09210-170, Brazil}

\begin{abstract}
In this letter we present a protocol to engineer interactions confined to
subspaces of the Fock space in trapped ions: we show how to engineer upper-,
lower-bounded and sliced Jaynes-Cummings (JC) and anti-Jaynes-Cummings (AJC)
Hamiltonians. The upper-bounded (lower-bounded) interaction acting upon Fock
subspaces ranging from $\left\vert 0\right\rangle $ to $\left\vert
M\right\rangle $ ($\left\vert N\right\rangle $ to$\ \infty$), and the sliced
one confined to Fock subspace ranging from $\left\vert M\right\rangle $ to
$\left\vert N\right\rangle $, whatever $M<N$. Whereas the upper-bounded JC or
AJC interactions is shown to drive any initial state to a steady Fock state
$\left\vert N\right\rangle $, the sliced one is shown to produce steady
superpositions of Fock states confined to the sliced subspace $\left\{
\left\vert N\right\rangle \text{,}\left\vert N+1\right\rangle \right\}  $.

\end{abstract}

\pacs{32.80.-t, 42.50.Ct, 42.50.Dv}
\maketitle

With the advent of quantum computation and communication, it has become
mandatory the development of techniques for the strict control of the coherent
manipulation of quantum states. Since the mid-1990s, much has been
accomplished with techniques for engineering effective Hamiltonian \cite{EH}
for preparation of non-classical states \cite{PNCS} and manipulation of their
evolutions. Concurrently, we have witnessed notorious progress on techniques
for controlling decoherence, with proposals like error correction \cite{EC},
decoherence-free subspaces \cite{DFS}, reservoir engineering
\cite{RE,CZ,Ruynet,Prado} and quantum feedback \cite{QF}, all already
implemented experimentally \cite{Exp, Haroche}.

More specifically, the search for steady Fock states has long been sought in
the framework of quantum computation and communication \cite{Haroche}, and has
led recently to a noticeable result in cavity QED \cite{Haroche}: the
generation of Fock states with photon number $n$ up to $7$ was driven, with
probability around $0.4$, by using a quantum feedback procedure to correct
decoherence-induced quantum jumps. Nonequilibrium number states up to $2$
photons were long before prepared in cavity QED \cite{Walther} as well as in
most suitable platforms as ion traps \cite{Leibfried} and, lately, in circuit
QED \cite{Cleland} where number states up to $6$ was achieved.

In the present letter, we bring to ion trap systems some of the
above-mentioned elements to produce high-fidelity quasi-steady motional Fock
states and steady superpositions of Fock states. The protocol proposed relies
on two key ingredients: the engineering of interactions lying in specific
subspaces of the Fock space and engineered reservoirs \cite{CZ,Ruynet,Prado}.
Specifically to the former, it is demonstrated how one can derive effective
Jaynes-Cummings ($JC$) and anti-Jaynes-Cummings ($AJC$) interactions which
confine a state evolution to the subspaces $\left\vert 0\right\rangle $ to
$\left\vert M\right\rangle $ or $\left\vert N\right\rangle $ to $\left\vert
\infty\right\rangle $. Therefore, the spectral decomposition of the time
evolution of such states will be upper and lower-bounded ($ub$ and $lb$) in
the Fock space. In addition, we also show a scheme for a slicing of the Fock
space, tailoring the system interactions in order to have them confined to
Fock subspaces ranging from $\left\vert M\right\rangle $ to $\left\vert
N\right\rangle $, with $M<N$.

The engineering reservoir technique is required to produce a Lindblad
absorption band, due to the $ub$ $AJC$ interaction, whose competition with the
natural Lindblad emission terms can be adjusted to favor the absorptive
process, thus leading any initial motional state to a quasi-steady Fock state
$\left\vert N\right\rangle $. Moreover, when the sliced interaction is used to
generate an equally sliced Lindblad superoperator acting upon the subspace
$\left\{  \left\vert N\right\rangle ,\left\vert N+1\right\rangle \right\}  $,
we show that the parameters can be conveniently adjusted to drive any initial
state to a given superposition $\left\vert \psi\right\rangle =c_{N}\left\vert
N\right\rangle +c_{N+1}\left\vert N+1\right\rangle $, thus extending the
scheme in Ref. \cite{Ruynet} which applies to the specific subspace $\left\{
\left\vert 0\right\rangle ,\left\vert 1\right\rangle \right\}  .$

We start with the Hamiltonian for the coupling between the electronic and
motional degrees of freedom of the trapped ion, the former described by the
raising $\sigma_{+}=\left\vert e\right\rangle \left\langle g\right\vert $ and
lowering $\sigma_{-}=\sigma_{+}^{\dagger}$ operators, $\left\vert
g\right\rangle $ and $\left\vert e\right\rangle $ standing for the ground and
excited states, respectively, and the latter described by the annihilation $a$
and creation $a^{\dagger}$ operators. In the resolved sideband regime ---where
the detuning between the electronic transition frequency $\omega_{0}$ and the
laser beam $\omega$ (used to couple the ionic degrees of freedom) is an
integer $k$ multiple of the vibrational frequency $\nu$, i.e. $k=\left(
\omega_{0}-\omega\right)  /\nu$---, we obtain
\[
H=\Omega\operatorname*{e}\nolimits^{i\phi-\eta^{2}/2}\sigma_{+}\sum
_{l=0}^{\infty}\frac{\left(  i\eta\right)  ^{2l+k}}{l!\left(  l+k\right)
!}\left(  a^{\dagger}\right)  ^{l}a^{l}a^{k}+H.c.\text{,}%
\]
$\Omega$ being the Rabi frequency, $\eta$ the Lamb-Dicke parameter and $\phi$
the phase of the laser field used to couple both ionic degrees of freedom. By
tuning the laser beam to the first red (blue) sideband and working to second
order in the Lamb-Dicke parameter, we derive the interaction%
\[
H_{k=\pm1}=\chi(\eta)\left[  \mathbf{A}(\eta)\sigma_{\pm}+\mathbf{A}^{\dagger
}(\eta)\sigma_{\mp}\right]  \text{,}%
\]
where $\chi(\eta)=\eta\left(  1-\eta^{2}/2\right)  \Omega$ and $\mathbf{A}%
(\eta)=\left[  1-\eta^{2}a^{\dagger}a/2\right]  a$. Expanding the operators
$A$ and $A^{\dagger}$ in the Fock space basis and adjusting the Lamb-Dicke
parameter to $\eta^{2}=2/N$ $\left[  \eta^{2}=2/\left(  N-1\right)  \right]  $
such that $\mathbf{A}^{\dagger}(\eta)\left\vert N\right\rangle =0$ $\left[
\mathbf{A}(\eta)\left\vert N\right\rangle =0\right]  $, we readily note that
$H_{k=\pm1}$ is decomposed in a sum of upper- ($ub$) and lower-bound ($lb$)
Hamiltonians, in the form
\begin{subequations}
\label{1}%
\begin{align}
H_{k=\pm1}  &  =H_{\pm}^{(ub)}+H_{\pm}^{(lb)}\text{,}\nonumber\\
H_{\pm}^{(ub)}  &  =%
{\displaystyle\sum\limits_{n=0}^{N-1}}
\chi_{n}\left(  \left\vert n\right\rangle \left\langle n+1\right\vert
\sigma_{\pm}+H.c.\right)  \text{,}\label{1a}\\
H_{\pm}^{(lb)}  &  =%
{\displaystyle\sum\limits_{n=N+1}^{\infty}}
\chi_{n}\left(  \left\vert n\right\rangle \left\langle n+1\right\vert
\sigma_{\pm}+H.c.\right)  \text{,} \label{1b}%
\end{align}
where $\chi_{n}=\sqrt{n+1}\left(  1-\eta^{2}n/2\right)  \chi(\eta)$ and the
($ub$ or $lb$) Hamiltonians $H_{+}$ and $H_{-}$ clearly stand for the JC and
AJC interactions, respectively. These interactions become incommunicable when
we prepared the vibrational state confined to the $ub$ or the $lb$ subspace,
$\left\vert 0\right\rangle $ to $\left\vert N\right\rangle $ or $\left\vert
N+1\right\rangle $ to $|\infty\rangle$: The evolution of any prepared state
$\rho=%
{\textstyle\sum\nolimits_{m,n=0}^{N}}
p_{mn}\left\vert m\right\rangle \left\langle n\right\vert $ or $\rho=%
{\textstyle\sum\nolimits_{m,n=N}^{\infty}}
p_{mn}\left\vert m\right\rangle \left\langle n\right\vert $, whatever the
electronic state is, remains indeed confined within the $ub$ or the $lb$
subspace. The considered second-order approximation holds for $N$ up to $10$
with typical parameters of ion trap experiments, allowing our technique to be
applied for several states of Fock space.

We have thus engineered JC or AJC interactions restricted to subspaces
$\left\vert 0\right\rangle $ to $\left\vert N\right\rangle $ or $\left\vert
N+1\right\rangle $ to $|\infty\rangle$, whatever the integer $N$ is, adjusted
through the choice of $\eta$. Additionally to that feature, one can envision a
use of the engineered interactions Eq. (\ref{1}) in order to decouple the
vibrational and internal atomic degrees of freedom. Indeed, by considering a
bicromatic field (generated through the above considered laser plus an electro
optical modulator), tuned to the first red and blue sidebands simultaneously,
one obtains $H=H^{(ub)}+H^{(lb)}$, where
\end{subequations}
\begin{subequations}
\label{2}%
\begin{align}
H^{(ub)}  &  =%
{\displaystyle\sum\limits_{n=0}^{N-1}}
\chi_{n}\left(  \left\vert n\right\rangle \left\langle n+1\right\vert
+H.c.\right)  \sigma_{x}\text{,}\label{2a}\\
H^{(lb)}  &  =%
{\displaystyle\sum\limits_{n=N+1}^{\infty}}
\chi_{n}\left(  \left\vert n\right\rangle \left\langle n+1\right\vert
+H.c.\right)  \sigma_{x}\text{.} \label{2b}%
\end{align}
Observe that a choice for the atomic state as an eigenstate of $\sigma_{x}$
effectively decouples the vibrational and internal degrees of freedom,
enabling one to directly select the $ub$ or $lb$ vibrational subspaces.

Now we turn our attention to engineer a Hamiltonian which confines the
dynamics of the vibrational state to a slice of the Fock space. For that, let
us consider again two laser beams. One of them electro-optically tuned to the
carrier ($\Omega_{1}$) as well as the first red ($\Omega_{3}$) and blue
($\Omega_{4}$) sidebands, while the other ($\Omega_{2}$) is tuned to resonance
with the electronic transition. Working again to second order in the
Lamb-Dicke parameters adjusted such that $\eta_{1}^{2}=\eta_{4}^{2}=2/\left(
N+1\right)  $, $\eta_{2}^{2}=2/N$, and $\eta_{3}^{2}=2/\left(  N-1\right)  $,
it follows the interaction%
\end{subequations}
\begin{equation}
H=\Omega\left(  \mathbf{B}\sigma_{+}+\mathbf{B}^{\dagger}\sigma_{-}\right)
\text{,} \label{3}%
\end{equation}
where $\mathbf{B}=\bar{\Omega}_{1}\mathbf{N}_{1}+\bar{\Omega}_{2}%
\mathbf{N}_{2}+\bar{\Omega}_{3}\mathbf{N}_{3}a+\bar{\Omega}_{4}a^{\dagger
}\mathbf{N}_{4}$ and $\mathbf{N}_{j}=\boldsymbol{1}-\eta_{j}^{2}a^{\dagger
}a/2$. We have also adjusted the Rabi frequencies to obtain $\bar{\Omega}%
_{1}=\Omega_{1}/\Omega=\left(  N+1\right)  \sqrt{N+1}/(N-1)$, $\bar{\Omega
}_{2}=\Omega_{2}/\Omega=N\sqrt{N+1}/(N+1)$, and $\bar{\Omega}_{3}=\Omega
_{3}/\Omega$, $\bar{\Omega}_{4}=\bar{\Omega}_{3}^{-1}$. It is straightforward
to verify that, for a prepared state $\left\vert \psi\right\rangle
=c_{N}\left\vert N\right\rangle +c_{N+1}\left\vert N+1\right\rangle $ with
$c_{N}/c_{N+1}=\bar{\Omega}_{3}$, the evolution governed by Hamiltonian
(\ref{3}) confines $\left\vert \psi\right\rangle $ to the subspace $\left\{
\left\vert N\right\rangle ,\left\vert N+1\right\rangle \right\}  $. Although
this Hamiltonian does not apply for $N=0$ or $1$ because of our choice of the
Lamb-Dicke parameters, the case $N=1$ can be implemented by considering
engineering interactions (confined to the subspace $\left\{  \left\vert
1\right\rangle ,\left\vert 2\right\rangle \right\}  $) using two laser beams,
each electro-optically tuned to two carrier transitions and the first blue
sideband. The first laser is set to be within the Lamb-Dicke regime, with the
phase adjusted to introduce a global phase factor $e^{i\pi}$ in all
transitions, while the second one has to be treated under a second order
approximation in appropriately adjunted Lamb-Dicke parameters. We finally
stress that, under the same considerations used to derive the interaction
(\ref{3}), we may engineer a Hamiltonian to confine the evolution of any
initial state $\rho=%
{\textstyle\sum\nolimits_{m,n=N}^{N+\ell}}
p_{mn}\left\vert m\right\rangle \left\langle n\right\vert $ to the subspace
$\left\{  \left\vert N\right\rangle ,...,\left\vert N+\ell\right\rangle
\right\}  $ using $\ell$ additional laser beams. For this purpose we set the
coefficients of the superposition $\left\vert \psi\right\rangle $ so that
$\mathbf{B}\left\vert \psi\right\rangle =\lambda\left\vert \psi\right\rangle
$, and as we will check below, with the additional condition $\lambda=0$ the
confined state is as well protected against decoherence.

Turning to the applications of the above derived (seven) Hamiltonians, we
first present a method to protect a Fock state which relies on engineered
reservoir, where an auxiliary decaying system (here the electronic levels of
the ion) is used to protect the state of the system of interest (the ionic
vibrational degrees of freedom). When considering the $ub$ $AJC$ interaction
$H_{-}^{(ub)}$, it is straightforward to obtain, by analogy with Refs.
\cite{CZ,Ruynet,Prado}, the engineered master equation%
\begin{equation}
\mathcal{L}_{eng}\rho=\frac{\Gamma}{2}\left(  2\mathbf{A}^{\dagger}%
\rho\mathbf{A-\mathbf{A}A}^{\dagger}\rho-\rho\mathbf{AA}^{\dagger}\right)  ,
\label{4}%
\end{equation}
with the effective damping rate $\Gamma=4\chi^{2}/\kappa$, $\kappa$ being the
damping rate of the internal excited state of the ion. As for interaction Eq.
(\ref{1b}), the action of this superoperator, derived by getting rid of the
electronic degrees of freedom, is confined to the $ub$ vibrational subspace.
Analyzing the complete equation $\dot{\rho}=\mathcal{L}_{eng}\rho
+\mathcal{L}\rho$, with $\mathcal{L}\rho=\left[  \left(  1+\bar{n}\right)
\gamma/2\right]  \left(  2a\rho a^{\dag}-\rho a^{\dag}a-a^{\dag}a\rho\right)
+\left(  \bar{n}\gamma/2\right)  \left(  2a^{\dag}\rho a-\rho aa^{\dag
}-aa^{\dag}\rho\right)  $ standing for the contribution of the thermal
environment, it is not difficult to conclude that under the condition
$\Gamma\gg\gamma$, any initial state $\rho=%
{\textstyle\sum\nolimits_{m,n=0}^{N}}
p_{mn}\left\vert m\right\rangle \left\langle n\right\vert $ is asymptotically
driven to a quasi-steady Fock state $\left\vert N=2/\eta^{2}\right\rangle $.
This occurs because the engineered contribution $\mathcal{L}_{eng}\rho$,
confined to the subspace $\left\vert 0\right\rangle $ to $\left\vert
N\right\rangle $, prevails over the action of the thermal environment. In Fig.
1, starting with the vibrational thermal state $\rho_{th}=%
{\textstyle\sum\nolimits_{n}}
\bar{n}^{n}/\left(  1+\bar{n}\right)  ^{1+n}$ $\left\vert n\right\rangle
\left\langle n\right\vert $ ($\bar{n}\approx0.01$ being the typical average
occupation number and $k_{B}$ being the Boltzmann constant) and adjusting
$\eta^{2}=2/M$, we present the evolutions of the fidelity $\mathcal{F}(t)=$
$\operatorname*{Tr}\left\vert M\right\rangle \left\langle M\right\vert
\rho(t)$ against $\gamma t$, considering typical vibrational decay rate
$\gamma_{0}\sim10$ Hz (where $\gamma=\gamma_{0}(1+M)^{0.7}$ \cite{Ruynet1}),
$\kappa\sim4\times10^{6}$ Hz, and $\Gamma\sim10^{4}\gamma$. As shown by the
black and grey dotted lines, the vibrational mode has been driven to steady
Fock states $M=5$ ($\eta^{2}=0.4$, $\Omega=1.2\times10^{6}$ Hz) and $10$
($\eta^{2}=0.2$, $\Omega=1.8\times10^{6}$ Hz), with significantly high
fidelities, around $0.98$, up to the relaxation time. In the inset of Fig. 1
we plot the Mandel $Q$-factor to inform us how close are the achieved steady
states of the desired Fock states $\left\vert 5\right\rangle $ and $\left\vert
10\right\rangle $, for which $Q=-1$. We verify that the steady state reached
with $M=5$ is significantly closer to a Fock state, showing $Q=-0.88$.
However, the state reached with $M=10$ shows $Q=-0.77$, a value that starts to
deviate significantly from the desired $Q=-1$ even though this state exhibits
a fidelity around that reached with $M=5$ and presents an unequivocally
sub-poissonian statistics.

We observe that, although\textbf{ }the interaction (\ref{1a}) is not suited
for state protection (since the thermal reservoir inevitably drives to the
vacuum any vibrational state initially confined to the $lb$ subspace), it is
perfectly suited, as well as all other interactions here engineered, for the
implementation of quantum-scissors device for optical state truncation
\cite{QS}.
\begin{figure}[ptb]
\includegraphics[width=0.8\textwidth]{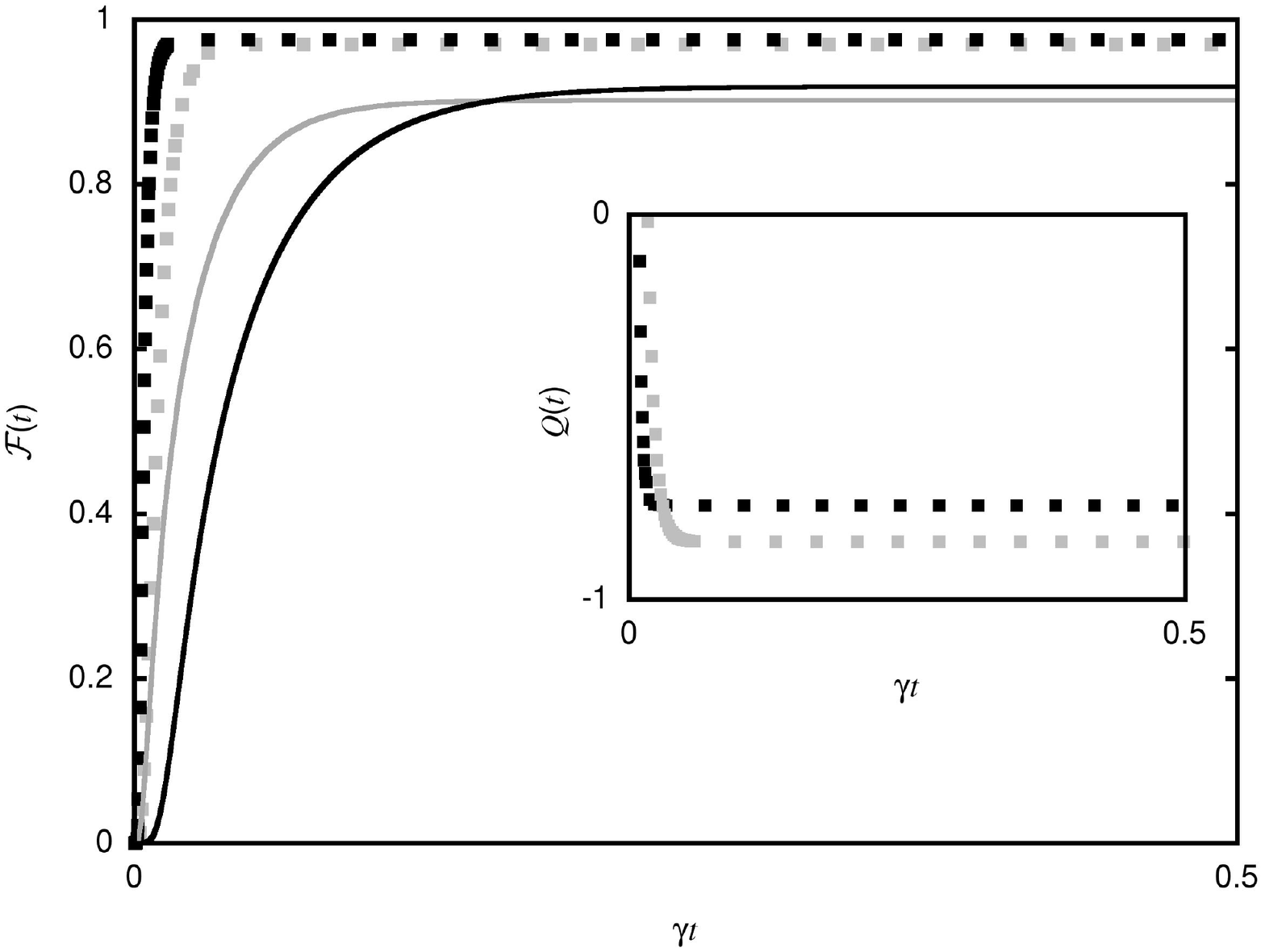}  \caption{Considering typical $\gamma_{0}\sim10$ Hz, $\kappa\sim4\times10^{6}$
Hz, and $\Gamma\sim10^{4}\gamma$, the black and grey dotted lines indicate the
evolutions of the fidelity $\mathcal{F}(t)=$ $\operatorname*{Tr}\left\vert
M\right\rangle \left\langle M\right\vert \rho(t)$ against $\gamma t$ for the
Fock states $M=5$ ($\eta^{2}=0.4$, $\Omega=1.2\times10^{6}$ Hz) and $10$
($\eta^{2}=0.2$, $\Omega=1.8\times10^{6}$ Hz), respectively, when starting
from the thermal state $\bar{n}\approx0.01$. Starting from the same thermal
state and $\tilde{\Gamma}\sim10^{4}\gamma$, the black and grey solid lines
indicate the evolutions of the fidelity $\mathcal{F}(t)=\operatorname*{Tr}%
\left\vert \psi\right\rangle \left\langle \psi\right\vert \rho(t)$ against
$\gamma t$, for the cases $M=4$ and $9$, for which we have adjusted
$\Omega\sim5.9\times10^{5}$Hz and $\Omega\sim7.3\times10^{5}$Hz, respectively.
In the inset we display the evolutions of the Mandel's $Q$ factor for the
generation of the Fock states $M=5$ and $10$}
\label{fig:fig1}%
\end{figure}
Using again the protocol for engineering reservoir \cite{CZ,Ruynet,Prado}, we
start from the Hamiltonian (\ref{3}) to obtain the engineered master equation%

\begin{equation}
\mathfrak{L}_{eng}\rho=\frac{\tilde{\Gamma}}{2}\left(  2\mathbf{B}%
\rho\mathbf{B}^{\dagger}\mathbf{-\mathbf{B}^{\dagger}B}\rho-\rho
\mathbf{B}^{\dagger}\mathbf{B}\right)  , \label{5}%
\end{equation}
with $\tilde{\Gamma}=4\Omega^{2}/\kappa$. Similarly to what happens with Eq.
(\ref{4}), the action of the Liouvillian (\ref{5}) is confined to the sliced
subspace $\left\{  \left\vert N\right\rangle ,\left\vert N+1\right\rangle
\right\}  $, with $N\neq0,1$, defined by the Hamiltonian from which it was
engineered. Turning now to the equation $\dot{\rho}=\mathfrak{L}_{eng}%
\rho+\mathcal{L}\rho$, we verify that under the condition $\tilde{\Gamma}%
\gg\gamma$ ---causing the contribution $\mathfrak{L}_{eng}\rho$ to prevail
over $\mathcal{L}\rho$--- any superposition $\left\vert \psi\right\rangle
=c_{M}\left\vert M\right\rangle +c_{M+1}\left\vert M+1\right\rangle $, where
$\mathbf{B}\left\vert \psi\right\rangle =0$, is protected against decoherence.
Starting again with the typical values $\bar{n}\approx0.01$ and $\tilde
{\Gamma}\sim10^{4}\gamma$, and adjusting $\eta_{1}^{2}=\eta_{4}^{2}=2/\left(
M+1\right)  $, $\eta_{2}^{2}=2/M$, $\eta_{3}^{2}=2/\left(  M-1\right)  $, we
also present in Fig. 1 the evolutions of the fidelity $\mathcal{F}%
(t)=\operatorname*{Tr}\left\vert \psi\right\rangle \left\langle \psi
\right\vert \rho(t)$ against $\gamma t$, for the cases $M=4$ and $9$, for
which we have adjusted $\Omega\sim5.9\times10^{5}$Hz and $\Omega\sim
7.3\times10^{5}$Hz, respectively. We verify that the vibrational mode has been
driven to the equilibrium superposition $\left\vert \psi\right\rangle $, with
exceptional high fidelity, around $0.90$, as shown by the black and grey solid
lines. We finally observe that while the protection of superposition states is
based on the protocol originally proposed in Ref. \cite{CZ} and adopted in
Ref. \cite{Ruynet}, our protocol for the protection of Fock states clearly
differs from that in Ref. \cite{CZ}. In fact, the protection of a
superposition state $\left\vert \psi\right\rangle $ demands the eigenvalue
equation $\mathbf{B}\left\vert \psi\right\rangle =0$ as required in Ref.
\cite{CZ}. However, although the condition $\mathbf{A}^{\dagger}%
(\eta)\left\vert N\right\rangle =0$ is automatically fulfilled to generate our
required $ub$ Hamiltonian, it is only a necessary condition. Our protocol for
the protection of Fock states also demands the dynamics to be confined within
the $ub$ subspace during the whole time evolution.

We have thus presented an original protocol to slice the Fock space, i.e., to
engineer upper-, lower-bounded and sliced JC and AJC Hamiltonians, which are
confined to subspaces of the Fock space. These Hamiltonians are used to
produce quasi-steady Fock states $\left\vert N\right\rangle $ and steady
superpositions of Fock states confined to the sliced subspaces $\left\{
\left\vert N\right\rangle \text{,}\left\vert N+1\right\rangle \right\}  $. Our
protocol can also be used for the implementation of quantum scissors, which
shows its suitability in the implementation of quantum logical devices and to
test the foundations of quantum mechanics.

\begin{flushleft}
{\Large \textbf{Acknowledgements}}
\end{flushleft}

The authors acknowledge the support from FAPESP, CNPQ and CAPES, Brazilian agencies.

\end{document}